%% file: main.tex
\documentclass[conference]{IEEEtran}

\usepackage{cite}
\usepackage{amsmath,amssymb,amsfonts}
\usepackage{graphicx}
\usepackage[hidelinks]{hyperref}
\usepackage{xcolor}
\usepackage[noend]{algpseudocode}
\usepackage{algorithm}

\definecolor{color_compute}{HTML}{5aa02c}
\definecolor{color_memory}{HTML}{d4aa00}
\definecolor{color_shim}{HTML}{2c89a0}
\definecolor{color_command}{HTML}{800000}

\begin{document}

\date{}

\title{Unlocking the AMD Neural Processing Unit for ML Training on the Client Using Bare-Metal-Programming Tools}

\author{
André Rösti\\
\textit{University of California, Irvine}\\
\textit{Irvine, CA, USA}\\
\textit{aroesti@uci.edu}
\and
Michael Franz\\
\textit{University of California, Irvine}\\
\textit{Irvine, CA, USA}\\
\textit{franz@uci.edu}
}

\maketitle{}

\input{abstract.tex}
\input{introduction.tex}
\input{related_work.tex}
\input{background.tex}
\input{design_overview.tex}
\input{design_integration.tex}
\input{design_npu.tex}
\input{evaluation.tex}
\input{future_work.tex}
\input{acknowledgments.tex}

\bibliographystyle{IEEEtran}
\bibliography{IEEEabrv,references}

\end{document}

%% file: abstract.tex
\begin{abstract}

There has been a growing interest in executing machine learning (ML) workloads on the client side for reasons of 
customizability, privacy, performance, and availability. In response,
hardware manufacturers have begun to incorporate so-called \textit{Neural Processing Units (NPUs)} into their processors for consumer devices.
Such dedicated hardware optimizes both power efficiency \textit{and} throughput for common machine learning tasks.
AMD's NPU, part of their Ryzen AI processors, 
is one of the first such accelerators integrated into a chip with an \textit{x86} processor.
AMD supports bare-metal programming of their NPU rather than limiting programmers to pre-configured libraries.

In this paper, we 
explore the potential of using a bare-metal toolchain to
accelerate the weight fine-tuning of a large language model,
GPT-2, entirely on the client side using the AMD NPU.
Fine-tuning on the edge allows for private customization of a model to a specific use case.
To the best of our knowledge, this is the first time such an accelerator has been used to perform training on the client side.
We offload time-intensive matrix multiplication operations from the CPU onto the NPU, achieving a speedup of over $2.8\times$ for these operations.
This improves end-to-end performance of the model in terms of throughput ($1.7\times$ and $1.2\times$ speedup in FLOPS/s on mains and battery power, respectively) and energy efficiency ($1.4\times$ improvement in FLOPS/Ws on battery power).
We detail our implementation approach and present an in-depth exploration of the NPU hardware and 
bare-metal tool-flow.

\end{abstract}

%% file: introduction.tex
\section{Introduction}

Ever-more powerful hardware has long been a principal enabler of breakthroughs in machine learning (ML);\cite{gpu-training}
stronger computational capabilities, i.e. \textit{FLOP/s} (floating point operations per second), along with greater data transfer bandwidths, enable more complex and better-trained models
within a given training time budget.
As we enter an era of personal computing where artificial intelligence (AI) applications such as chatbots
pervade many workflows,
there is an increased interest in running these applications entirely on the client side for customizability, privacy, performance, and availability.
While most such applications at the edge only perform inference, customization also requires training (fine-tuning) models.
Given that edge devices, such as laptops, often run on battery power, such use cases demand excellent energy efficiency.
More \textit{FLOP/W} (floating point operations per Watt) enable running or training larger parts of a model on the end-user's device without compromising on battery life.

While graphics processing units (GPUs) have evolved to meet the ever-increasing demand for computation capabilities (\textit{FLOP/s}), 
their power consumption is generally high.
Recognizing shifting demands, various manufacturers have begun shipping dedicated \textit{neural processing units (NPUs)} as part of their processors.
These systems aim to strike a new balance between throughput and power consumption, optimizing \textit{FLOP/Ws} (floating point operations per Watt-second) jointly.
Among these, \textit{AMD}'s \textit{Ryzen AI} is the first system-on-a-chip incorporating a dedicated NPU alongside an \textit{x86} CPU and integrated GPU.
The AMD NPU's architecture, called \textit{XDNA}, is comprised of a spatial array of so-called \textit{AI Engines}, each of which can independently perform computation.

AMD provides multiple compiler tool-flows to program the NPU. 
Similar to competitor's offerings, AMD's principal production tool-flow, called \textit{Ryzen AI Software}, 
provides a library of highly optimized computation kernels that run on the NPU.
A lower-level tool-flow, \textit{IRON}\cite{online-iron}, enables bare-bones hardware access using an \textit{MLIR} dialect and Python bindings.
Development at this level is more labor intensive, but allows programmers to harness every hardware feature and tailor their application implementation for the device.

In this paper, we explore the use of the AMD NPU and the \textit{IRON} tool-flow to accelerate client-side inference \textit{and fine-tuning (training)} of a large language model, GPT-2.
GPT-2 
is a good example for the kind of workload that end-users may want to run and train locally;
fine-tuning it enables a customized user experience without compromising privacy.
We demonstrate that this is feasible using current hardware and tool-flows,
by implementing a specialized matrix-multiplication kernel for the NPU that is tailored to the requirements of this application,
and then modifying a framework-free implementation of GPT-2, called \textit{llm.c}\cite{llm-c}, to efficiently use this kernel.
Our carefully obtained measurements show that our approach significantly improves throughput and power efficiency over the original implementation.

%% file: related_work.tex
\section{Related Work}

The rise of GPUs has made it possible to train larger ML models\cite{gpu-training,imagenet,deep-learning-overview}, waking us from the ``AI winter'' of the 1990s.
Today, due to the waning of Moore's law\cite{moores-end}, we must rely on hardware \textit{specialization}, i.e., dedicating circuitry to common domain-specific functions, for more processing power.
Google's tensor processing unit (TPU)\cite{google-tpu}, a matrix multiplication accelerator card, 
is an early example of such specialized hardware.
Both research\cite{diannao,cnn-fpga,cnn-fpga-exploration} and industry\cite{microsoft-isca-2014,microsoft-isca-2018} frequently use field-programmable gate arrays (FPGAs) to explore specialized neural network hardware designs.
Accelerating generalized matrix-matrix multiplication (GEMM) workloads\cite{sigma} is particularly interesting, as it is at the core of many ML applications.
AMD's NPU\cite{online-xdna,online-aie-whitepaper,ryzen-micro} and competitor products, like Apple's Neural Engine\cite{online-apple} or Qualcomm's NPU\cite{online-qualcomm}, are further instances of hardware specialization.
The AMD NPU's \textit{XDNA} architecture\cite{ryzen-micro} is more flexible than application-specific integrated circuits (ASICS) and more performant than general-purpose CPUs.
Other works have explored different points along this flexibility-performance trade-off, for example 
in coarse-grained reconfigurable architectures (CGRAs)\cite{cgra}.

The task of optimally mapping the work of an application onto an accelerator
is non-trivial\cite{gpu-mapping-course,gpu-mapping-streaming}.
GEMM algorithms have received particular attention\cite{kblas,stream-k} in the application mapping sphere;
most approaches tile the input matrices into submatrices to parallelize work.
Parameters of the design space, like tile size, can be automatically evaluated in a process called auto-tuning\cite{autotuning}.
Prior work has also explored mapping GEMM algorithms\cite{charm,charm2,maxeva} onto the AI Engines in AMD Versal devices; these findings also translate to the NPU, which is equipped with the same AI Engines.

This paper focuses on mapping the GPT-2\cite{gpt-2} large language model (LLM) onto the AMD NPU.
GPT-2, along with its successor\cite{gpt-3} and competitor\cite{llama} models, follows a transformer architecture\cite{attention}.

To implement these ML models, programmers traditionally have used frameworks\cite{ml-framework-survey,ml-edge-survey} such as PyTorch and TensorFlow, 
which provide the most common data structures (e.g., computation graphs) and algorithms (such as gradient descent).
The generality of these frameworks can add overheads and preclude specializations needed for optimal performance.
Because of this, programmers have begun to move towards using minimal, purpose-built ``disposable'' frameworks\cite{unstoppable-rise}.
Examples of this include \textit{llama2.c}\cite{llama-2-c}, an implementation of LLaMA 2 inference in one file of pure C code,
\textit{llama.cpp}\cite{llama-cpp}, which is the primary testing ground of the GGML framework,
and \textit{llm.c}\cite{llm-c}, the subject of this paper.
Previous work has ported \textit{llama.cpp} onto the AMD NPU\cite{online-ryzenai-llm-cpp} 
but 
performed inference only (no training).
We believe we are the first to map a training workload (fine-tuning) of a similar model (GPT-2) onto the AMD NPU.

%% file: background.tex
\section{Background}

This paper combines three ingredients:
(1) powerful \textit{hardware} (AMD's NPU), 
(2) an interesting \textit{application} (GPT-2), and
(3) a productive \textit{tool-flow} (IRON) that allows us to port our application onto the hardware.

\subsection{The Hardware: AMD XDNA NPU in Ryzen AI}

\begin{figure}
    \includegraphics{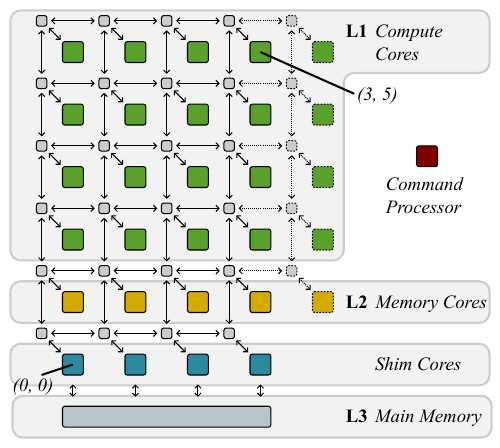}
    \caption{
        \label{fig:architecture-diagram}
        Overview of the XDNA architecture, showing the VLIW processor {\color{color_compute}compute cores (L1)}, also called AI Engines, {\color{color_memory}memory cores (L2)}, {\color{color_shim}shim cores} and the dedicated {\color{color_command}command processor}.
        The small grey boxes between arrows are configurable interconnect switch boxes.
    }
\end{figure}

\textit{XDNA} and its successor \textit{XDNA 2} are spatial computing architectures that arrange compute cores in a two-dimensional grid.
This paper focuses on the first iteration of the first-generation \textit{XDNA} architecture codenamed \textit{Phoenix}.
On this NPU, cores are arranged as a grid of four rows and five columns.
Four of these columns have a shim core that allows directly interfacing with main memory. 
Cores in the fifth column requiring access to main memory must route their requests via the shims in the first four columns.
For simplicity and to enable regularity in our design, we will focus on the $4\times4$ partition of the main grid that has shim cores throughout the rest of this paper.
We will identify each core by its zero-indexed $(x, y)$ coordinates from the bottom left.

Figure \ref{fig:architecture-diagram}, shows the three types of cores 
({\color{color_compute} compute cores}, {\color{color_memory}memory cores}, and {\color{color_shim}shim cores}) and their arrangement.
\footnote{AMD calls these ``compute \textit{tiles}'', ``memory \textit{tiles}'' and ``shim \textit{tiles}''. We use the term ``core'' to avoid confusion with the matrix \textit{tiling} discussed later.}
Each compute core has $64 \text{KB}$ of local memory and can execute code in parallel to the other cores.
Memory cores each add another $512 \text{KB}$ of memory and enable data reuse and distribution.
The shim cores enable moving data from the CPU and GPU in and out of the accelerator. 
A dedicated {\color{color_command}command processor} with access to all cores and switch boxes can be used to reconfigure the NPU at runtime.

To program the NPU, the programmer loads computation kernels onto the AI Engine cores and configures the data movement between cores and the CPU. 

\subsubsection{Data Movement}

The grey boxes in figure \ref{fig:architecture-diagram} label the memory levels (L1, L2, L3) of all cores.
L1 and L2 memories are local to cores; L3 corresponds to the unified main memory shared between the CPU, NPU, and integrated GPU, which can be directly accessed by each of these components.
Cores at each level have a number of data movement accelerators (DMAs), which are simple processors that can copy data to and from the interconnect and local core memories, and acquire and release hardware semaphore locks for synchronization.
A unique feature of the XDNA architecture is that the programmer must explicitly describe all data movement.
This is done by (i) setting up circuit- or packet-switched routes between the cores (streams) through the switch boxes and (ii) programming the DMAs of each core.
DMAs are independent of compute cores, so data movement can occur in parallel with computation.

\subsubsection{Compute Cores}

Compute cores (``AI Engines'') are very large instruction word (VLIW) processors with parallel issue slots for matrix multiplication operations, vector addition operations, vector shuffling and shifting, two slots for loading from memory and one slot for storing to memory.

There are no caches present in this hardware, and instructions have fixed latencies.
The hardware has minimal stalling infrastructure, and the compiler must schedule instructions to avoid hazards.

Due to the lack of stalling, back-to-back vector operations (without compiler-inserted no-ops) in the generated assembly directly indicate $100\%$ vector hardware utilization.

The vector processing units in the XDNA architecture can perform 128
fused-multiply add (FMA) operations 
for the \textit{bfloat16} input type and \textit{float32} output type
in every clock cycle\cite{online-versal-docs}.
At a 1 GHz clock frequency, this equals 256 GFLOP/s per AI Engine core,
for a total of 4 TFLOP/s of \textit{bfloat16} processing power for the $4\times 4$ partition of compute cores this paper targets.
When processing \textit{int8} data types, AMD advertises up to 10 TOP/s for the Phoenix-generation XDNA NPU. 

\subsection{The Application: GPT-2}

\begin{figure}
    \begin{center}
        \textbf{GPT-2 (124M) Floating Point Operations}
    \end{center}
    \begin{center}
        \includegraphics{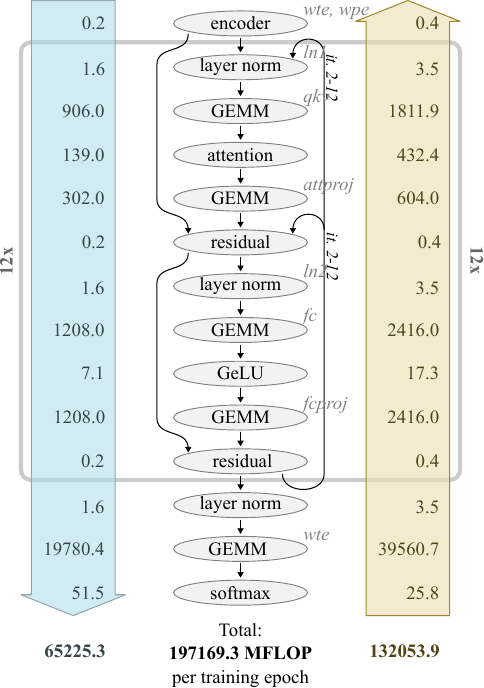}
    \end{center}
    \caption{
        \label{fig:operation-count}
        Computation graph of GPT-2, including floating point operation count ($1 \text{MFLOP} = 1e6 \text{FLOP}$).
        The fine-tunable weights (as named in \textit{llm.c}) are annotated in grey italics.
        The arrows on the left and right show the FLOP count in the forward and backward pass, respectively.
        The operations within the grey box are repeated twelve times.
    }
\end{figure}

Figure \ref{fig:operation-count} shows the computation graph of the 124M parameter variant of the GPT-2 large language model (GPT-2 small).
GPT-2 gained widespread recognition for being one of the first models to produce plausible responses to question-answering, machine translation, reading comprehension, and summarization prompts.
Even though more recent releases have far surpassed GPT-2's performance,
the architecture of newer large language models has remained largely identical.
The model's small size enables training locally on end-user devices, making it an interesting case study for us.

In the process of porting GPT-2 to the AMD NPU, we will focus our attention on offloading general matrix-matrix multiplication (GEMM) of the form $AB = C,$
where the $A$ matrix has $M$ rows and $K$ columns,
the $B$ matrix has $K$ rows and $N$ columns,
and the resulting output $C$ is of size $M\times N$.
We refer to these dimensions as the ``problem size'' and denote it as $M\times K \times N$.

\subsection{The tool-flow: IRON}

\begin{figure}
    \includegraphics{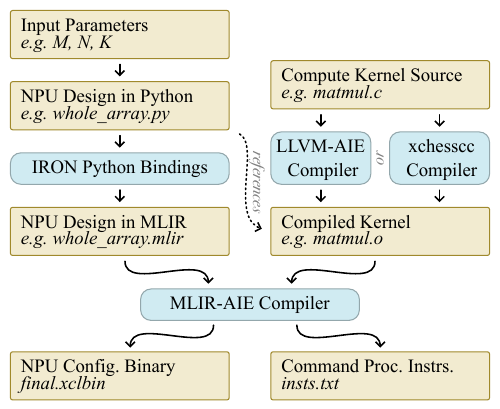}
    \caption{
        \label{fig:toolflow-diagram}
        Available tools (blue) and the intermediate outputs (brown) they produce in the open-source \textit{IRON} tool-flow
    }
\end{figure}

One of AMD's primary tool-flows to program the Ryzen AI processors is called \textit{Ryzen AI Software}\cite{online-ryzenai-software};
Using it, frameworks like PyTorch and TensorFlow can automatically offload some common operations in models with minimal programmer effort.
A similar tool-flow exists for AMD's accelerator and FPGA cards (Alveo, Versal), \textit{Vitis AI Software}\cite{online-vitisai-software}.
Alternatively, a \textit{Vitis}-based flow allows programming the Versal AIE devices at a lower level using C/C++ code.

Figure \ref{fig:toolflow-diagram} shows the components of \textit{IRON}\cite{online-iron}, an open-source tool-flow that allows low-level programming of AMD's NPU.
In \textit{IRON}, programmers use a Python script to describe the layout of their NPU design;
the data movement can either be described by manually configuring DMAs, switch boxes, and locks,
or via a higher-level \textit{ObjectFIFO} abstraction.
The script also specifies what code each compute core should execute
by reference to an object file compiled from C++ source code.
The C++ source code uses the \textit{AIE API}\cite{online-aie-api} and can be compiled with a choice of either an LLVM-based open-source compiler, or AMD's proprietary \textit{xchessccc} compiler.
Running the Python script generates an intermediate representation (IR) in MLIR\cite{lattner2021mlir} of the design for the MLIR-AIE compiler.
Compiling this IR results in two outputs:
A \textit{final.xclbin} file, containing the static configuration of all cores and switch boxes of the NPU,
and an \textit{insts.txt} file, which contains instructions to be executed by the dedicated command processor for re-configuring the NPU at runtime.

%% file: design_overview.tex
\section{Design: Overview}

We considered two approaches for our design:
(a) an end-to-end implementation of the entire model on the NPU (``data-flow''), or
(b) a hybrid NPU-CPU implementation, which runs only the most compute-intensive operations on the NPU and leaves all other operations to be performed on the CPU (``layer-by-layer'').
We choose to follow the more straightforward layer-by-layer approach, which allows incrementally offloading parts of a model.

We base our implementation on \textit{llm.c}\cite{llm-c},
an optimized implementation of GPT-2 written in pure C without external libraries.
Both \textit{llm.c} and our bare-metal NPU implementation forgo the use of general-purpose frameworks.
Since abstractions rarely are free, both \textit{llm.c}'s and our low-level approach promises maximized hardware utilization.

Profiling the baseline implementation reveals matrix multiplication as the main bottleneck, both in the forward and backward passes
(see ``CPU'' in figure \ref{fig:op-breakdown} in the evaluation section). 
We therefore choose to offload this operation onto the NPU.
To this end, we devise a parametrized design that can be used to generate NPU programs for multiple matrix sizes at build time, while being tailored to the sizes used in the 124M-parameter GPT-2 model at runtime.

%% file: design_integration.tex
\section{Design: CPU Side}

For the CPU side of our implementation, we modify the original \textit{llm.c},
replacing matrix multiplications with an invocation of our NPU design.

A challenge of the layer-by-layer approach is that it requires reconfiguring the accelerator between invocations of different kernels. We have observed that this reconfiguration can be a major contributor to overall overheads, especially for small matrix sizes where it is not amortized as well. Therefore, we make every effort to minimize the required NPU reconfiguration.

\subsection{Initialization}

The Xilinx Run Time (XRT) is the host programming interface that allows us to interface with the NPU during execution. 
At the beginning of our program, 
we use XRT to initialize the program memories of each AI Engine core
and configure the interconnect switch boxes of the L1 and L2 memory levels. 
To ensure this configuration occurs only once,
we designed our NPU kernels so that the data transfer always remains the same at these levels, irrespective of problem size.
We then pre-load one instruction stream for the NPU command processor per problem size. 
These instruction streams encode the data movement between L3 and L2. 
The instruction streams are generated and compiled at build time.
Lastly, we initialize shared XRT buffers to pass input and output matrices between CPU and NPU.
We allocate one set of shared buffers for each problem size we aim to support.

The result of initialization is a partially initialized NPU (level L2 and up) 
and a hash map that stores the XRT data structures (instruction streams, shared XRT buffers) for each problem size for later use.

\subsection{Matrix Multiplication Invocation}

We start each invocation of an offloadable matrix multiplication
by copying the input buffers to the corresponding shared XRT buffers.
This is necessary for our minimally-invasive modular implementation, which works irrespective of the call site; 
zero-copy buffers could be implemented by replacing the buffers used throughout the original implementation with shared XRT buffers.

In the original \textit{llm.c} implementation,
weights are stored in \textit{column-major} order, whereas activations are laid out in a \textit{row-major} order.
This leads to inconsistent data layouts across invocations when the derivatives are calculated during backpropagation.
Since our NPU design always expects the same data layout, 
we added code to additionally perform a transpose on the CPU as we copy the input buffers where needed.
We optimized this transpose by parallelizing it across all available CPU cores.

We could alternatively use the data layout transformation features of the DMAs in the NPU to perform this transpose;
however, this would require reconfiguration of nearly all DMAs between invocations, which is impractically slow. 
Yet another approach might be to rewrite \textit{llm.c} entirely to use all row-major data structures; however, this would significantly affect the cache locality, and hence the performance, of the CPU algorithms that we do not currently offload.

Once the input buffers are set up, we issue the instruction stream for the given problem size to the command processor.
This configures the L3 to L2 DMAs, which immediately start tiling data from the shared input buffers.
The instruction stream also contains instructions to write two runtime parameters into the registers of the AI Engine cores: 
the number of matrix tiles to accumulate and the number of tiles in the output matrix.

The CPU then waits for NPU execution to complete;
this happens as soon as the last L3 shim has copied the last output tile back into the shared output buffer.
The CPU then copies the results out of the shared output buffer into the application.

Note that this design is carefully engineered to minimize NPU reconfiguration (only L3 is reconfigured), which is critical to attaining workable performance.

%% file: design_npu.tex
\section{Design: NPU Side}
\label{sec:design-npu}

At a high level, the NPU side of our design tiles the input matrices $A$ (of size $M\times K$) and $B$ (of size $K \times N$),
which it reads from shared buffers in L3,
into sub-matrices of size $m\times k$ and $k \times n$, respectively, on the L2 memory cores.
These cores then repeatedly distribute tiles to the 16 computation cores (L1), each of which multiplies and accumulates its inputs in-place.
Memory cores (L2) join the output tiles produced by the computation cores (L1) and route them back to an output buffer in the main memory (L3).

The sizes of the input matrices and sub-matrix tiles are compile-time parameters of our design.
We generate one variation of our design for each of the 12 differently-sized matrix multiplications in 124M GPT-2 (listed in figure \ref{fig:mm-sizes}).
We tailor our design to the requirements of the application by using the tile dimensions $m=64$, $k=64$, $n=32$.
With these sizes, we only need to pad one input matrix of size $50304\times256$ to $50432\times256$.
All other matrix sizes are evenly divisible by our tile size, and we maximize usage of the available compute core memory.

\subsection{Computation Core Kernel}

Our design uses a $4\times 4$ partition of the computation cores of the NPU.
Each core runs the same code in parallel on different data;
specifically, each core executes a vectorized kernel that multiplies two sub-matrices $A'$ (of size $m\times k$) and $B'$ (of size $k\times n$), accumulating results into an output tile $C'$ of size $m \times n$.
To do so, the kernel initially zeroes $C'$,
and then repeatedly acquires two input tiles $A'$ and $B'$, multiplies them, and accumulates the result in-place into $C'$.
After $\frac{K}{k}$ such multiplications, the core's DMA sends the output tile back to the L2 memory cores.
This process is repeated for every output tile.

We use double-buffering for all buffers; that is, there are two physical buffers reserved to hold the $A'$, $B'$ and $C'$ tiles, and the DMA and computation core alternate between them.
While the DMA receives inputs and/or sends results from one set of buffers,
the computation core simultaneously performs its computation using previously received tiles stored in the other set of buffers.

Our computation kernel uses an AI Engine fused-multiply-add (FMA) instruction called \texttt{VMAC}.
This instruction multiplies two matrices of size $4 \times 8$ and $8 \times 4$ and adds the result to an accumulator register holding an output tile of size $4 \times 4$.
The result of this operation is available four cycles after the instruction was issued.
If a subsequent \texttt{VMAC} instruction uses the same accumulator register, the compiler must delay it by inserting no-ops to avoid a read-after-write data hazard.
To avoid such no-ops, we structure our kernel so that it simultaneously calculates \textit{four independent} output tiles, held in four distinct accumulator registers.
The inner-most loop of our kernel contains four independent $\texttt{VMAC}$ instructions that execute back-to-back.

To use \texttt{VMAC}, sub-matrices of sizes $4\times 8$ (for $A'$) and $8\times 4$ (for $B'$) must be laid out contiguously in memory.
We use both the computation core's DMA and the AI Engine's data swizzling instructions to ensure all data is tiled correctly.
First, the DMA swizzles data at a 4-byte granularity (the finest granularity it supports),
and then a so-called \texttt{VSHUFFLE} instruction swaps the remaining misplaced two bytes during execution.
\texttt{VMAC} and \texttt{VSHUFFLE} can simultaneously execute because they are implemented in separate hardware units.
As a result, these additional operations 
have no impact on runtime.
We verified this empirically by turning off the transformations (breaking the correctness of results) and observing identical runtimes.
Similarly, the \texttt{VLOAD} instructions required to load data into registers can execute in parallel,
so we can ensure the inner loop of our kernel is compute-bound.

By checking the generated assembly for back-to-back vector instructions (no no-ops for stalling), we were able to verify that the innermost loop of our kernel fully utilizes the vector hardware.
However, some pre- and postamble code (``filling the pipeline''), where utilization is not optimal, is necessary before entering such a loop.
By maximizing the tile size, we can minimize the number of these less efficient pre- and postambles.

\subsection{Tiling and Data Movement}

\begin{figure}
    \includegraphics{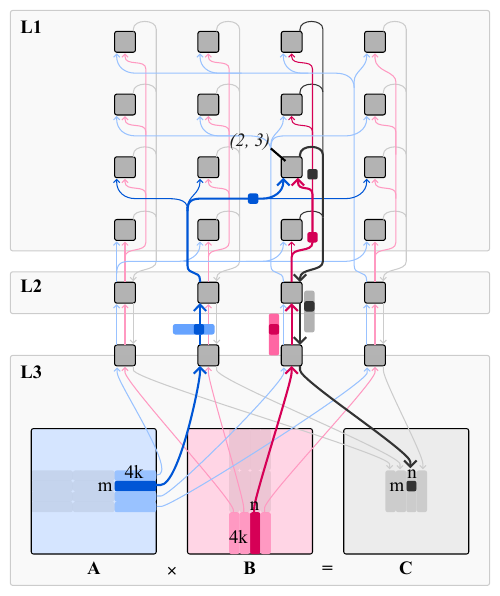}
    \caption{
        \label{fig:data-movement-diagram}
        Data movement and tiling of input and output matrices across the three memory levels.
        The data movement to the compute core \textit{(2, 3)} is highlighted as an example:
        This core receives its sub-tile of the \textit{A} matrix from the memory core in column 1,
        and its sub-tile of the \textit{B} matrix from the memory core in column 2 (zero-indexed).
    }
\end{figure}

Figure \ref{fig:data-movement-diagram} shows the data movement between the NPU cores and the host's main memory.
Tiling the input matrix into sub-matrices enables parallelism between compute cores, as well as overlapped data movement and computation. 
The limited memory in the compute cores also makes tiling a necessity, as
the entirety of the matrices does not fit into a compute core's L1 memory.

We are following an \textit{accumulate-in-place} tiling recipe;
our implementation iterates through the $m\times n$-sized output tiles of the output matrix $C$ in-order.
We stream the required input tiles into the cores, and each compute core locally accumulates an entire output tile before streaming it back out.

More specifically, the L3 shim cores stream $m\times k$-sized and $k\times n$-sized sub-tiles of the input matrices A and B, respectively, into the L2 memory cores.
Each of the four shim cores is responsible for one quarter of the input space.
Rows of tiles of $A$ are repeated $\frac{N}{4n}$ times, and columns of tiles of $B$ are repeated $\frac{M}{4m}$ times.
Transfers on each shim core are offset by a multiple of the tile size and its (zero-based) hardware column index.

That is, the shim core in the hardware column $\mathbf{i} \in \left\{0, 1, 2, 3\right\}$ will transfer
input matrix $A$'s rows $\mathbf{i}m + 4jm$ through $\mathbf{i}m + 4(j+1)m-1$, for $j=0,1,2,\dots,\frac{M}{4m}$, tiled into $k$-column-wide blocks, 
and input matrix $B$'s columns $\mathbf{i}n + 4jn$ through $\mathbf{i}n + 4(j+1)n-1$, for $j=0,1,2,\dots,\frac{N}{4n}$, tiled into $k$-rows-tall blocks.

For example, the shim in the hardware column $3$ streams in the following sub-matrices of $A$, in sequence:
\small
\begin{gather*}
\begin{bmatrix}
A_{3m,0} & \hdots & A_{3m,k-1} \\
\vdots & \ddots & \vdots \\
A_{4m-1,0} & \hdots & A_{4m-1,k-1}
\end{bmatrix},
\begin{bmatrix}
A_{3m,k} & \hdots & A_{3m,2k-1} \\
\vdots & \ddots & \vdots \\
A_{4m-1,k} & \hdots & A_{4m-1,2k-1}
\end{bmatrix},\\
\dots,\\
\begin{bmatrix}
A_{3m,K-k-1} & \hdots & A_{3m,K-1} \\
\vdots & \ddots & \vdots \\
A_{4m-1,K-k-1} & \hdots & A_{4m-1,K-1}
\end{bmatrix},
\left(\text{repeat $\frac{N}{4n}$ times}\right),\\
\begin{bmatrix}
A_{7m,0} & \hdots & A_{7m,k-1} \\
\vdots & \ddots & \vdots \\
A_{8m-1,0} & \hdots & A_{8m-1,k-1}
\end{bmatrix},
\begin{bmatrix}
A_{7m,k} & \hdots & A_{7m,2k-1} \\
\vdots & \ddots & \vdots \\
A_{8m-1,k} & \hdots & A_{8m-1,2k-1}
\end{bmatrix},\\
\dots
\end{gather*}%
\normalsize

The L2 memory cores store blocks of four tiles of $A$ and $B$ locally (i.e. $m\times 4k$-sized and $4k\times n$-sized blocks).
These are then \textit{distributed} across compute cores as $m\times k$- and $k\times n$-sized tiles:
A memory core in the hardware column $i$ distributes input $A$'s tiles across all the compute cores in the hardware row $i+2$, such that core $(i+2, 0)$ receives the first tile, core $(i+2, 1)$ receives the second tile, and so on.
The input matrix $B$ is distributed across compute cores in the same hardware column $i$ as the memory core; 
that is, compute core $(2, i)$ receives the first tile, core $(3, i)$ receives the second tile, and so on.
\footnote{Note that row $2$ is the lowest row of compute cores.}

For every $\frac{K}{k}$ input tiles of $A$ and $B$ streamed into a compute core,
that core will produce one $m\times n$-sized output tile of $C$.
The L2 memory cores then perform a column-wise \textit{join} into $m\times 4n$-sized output tiles,
which are then written back into their appropriate positions in the $C$ output buffer by an L3 core.

The entirety of this data movement is 
parametrized by problem size $M$, $K$ and $N$, as well as tile size $m$, $k$, $n$.
This allows us to generate concrete design variants for different parameters.
The intricacy of this design shows the versatility of the tool-flow and the specialization possibilities that bare-metal hardware access enables.

\subsection{Data Layout Transformations}
\label{sec:data-layout-transformations}

\begin{figure}
    \includegraphics{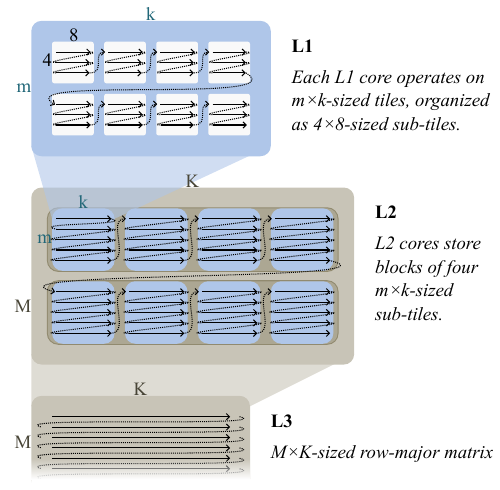}
    \caption{
        \label{fig:tiling-diagram}
        Data layout transformations performed on the hardware DMAs for matrix sub-tiles for input matrix A. Matrices B and C are analogously tiled with their respective dimensions.
    }
\end{figure}

The DMAs in the XDNA architecture can be configured to access data in non-linear patterns at a granularity of 4 bytes.
Figure \ref{fig:tiling-diagram} shows how we use this hardware feature to transform the layouts of sub-matrices of the input matrix $A$.
Concretely, we first transform matrix $A$, which resides in row-major format in L3, 
such that sub-matrices of size $m\times k$ are laid out contiguously in L2.
Going from L2 to L1, we then rearrange these $m\times k$-sized tiles to satisfy the vector intrinsic size requirements ($4\times 8$-sized tiles) in L1.
Analogously, the input matrix $B$ is transformed from a column-major layout in L3 into $k\times n$-sized tiles in L2, then $8\times 4$-sized tiles in L1.
Output $C$ morphs from $4\times 4$-sized tiles in L1 into $m\times n$-sized tiles in L2, and is then written out as a row-major matrix to L3.

\subsection{Reconfiguration}
\label{sec:npu-reconf}

By using the same tile size $m$, $k$, and $n$ for all variations, we
completely eliminate the need to reconfigure the compute (L1) and memory (L2) cores.
Only the shim cores and two runtime parameters in each core require reconfiguration between different problem sizes.
This minimal reconfiguration reduces the switching times between GEMM sizes.

The compute cores read two runtime parameters from memory:
The number of tiles to accumulate $\frac{K}{k}$ for each output tile, and the number of output tiles $\frac{MN}{mn}$ to produce before re-reading the parameters.
This allows the cores to adapt to new sizes after each complete GEMM.
When reconfiguring for different problem sizes, the command processor writes new values for these parameters to each core's memory.

%% file: evaluation.tex
\section{Evaluation}
\label{sec:evaluation}

We evaluated our implementation on a Asus Vivobook Pro 15 with an AMD Ryzen 9 7940HS CPU, equipped with a \textit{Phoenix} XDNA NPU.
We ran Ubuntu 22.04 LTS (kernel 6.10), and XRT/XDNA driver version 2.18.0.
We disabled dynamic frequency scaling on the CPU and ran all benchmarks in a non-GUI environment.
We restarted the test machine between each run.
We observed that an unloaded CPU is critical for a competitive CPU-only implementation.
In real-world use cases, where there may be concurrently running applications, this gives an additional edge to the NPU implementation. 

We timed 41 training epochs of GPT-2 individually (\textit{llm.c}'s default).
Each epoch consists of 197 GFLOP (see figure \ref{fig:operation-count}).
We measured power consumption by polling a power driver file (\textit{/sys/class/power\_supply/BAT0/power\_now}) every $\frac{1}{4} s$.

For benchmarks with high variance, the figures include boxes-and-whiskers to show the spread of the measurements; the standard deviation for all other figures is below $5\%$.

\subsection{GEMM Performance}

\begin{figure}
    \begin{center}
        \textbf{GEMM Performance}
    \end{center}
    \begin{center}
        Forward Pass Problem Sizes
    \end{center}
    \includegraphics{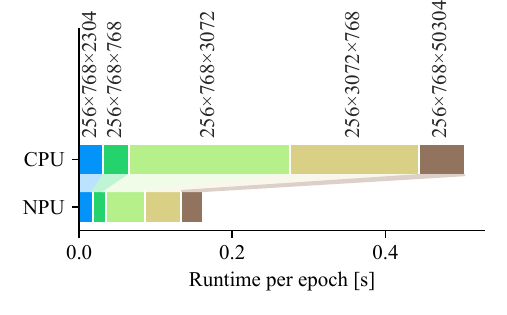}
    \begin{center}
        Backward Pass Problem Sizes
    \end{center}
    \includegraphics{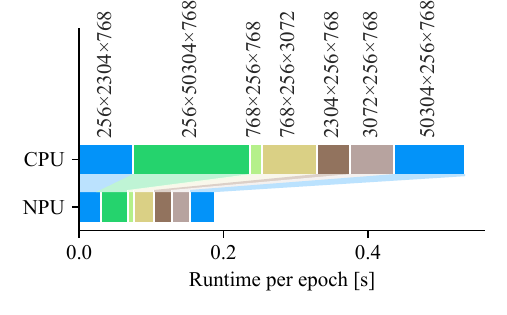}
    \caption{
        \label{fig:mm-sizes}
        Total runtime spent performing GEMM operations in each training epoch, split by problem size (mean of 41 training epochs, lower is better). Note that some GEMM sizes occur repeatedly in a single training epoch (in which case we show the sum of total runtime across all invocations), and the GEMM sizes issued in the forward pass also occur in the backward gradient calculations.
    }
\end{figure}

Figure \ref{fig:mm-sizes} visualizes the performance of our matrix multiplication design in isolation.
We observe that the NPU implementation is faster than the CPU for every problem size.
On average, GEMMs for sizes of the forward and backward passes are \textbf{$\mathbf{3.1\times}$ and $\mathbf{2.8\times}$ faster}, respectively.
GEMM size $256\times 50304 \times 768$ experienced the largest speedup of $4.2\times$, whereas the size $256\times 768\times 2304$ improved the least with a $1.8\times$ speedup.
In general, the relative speedup is largest for the larger problem sizes, because larger sizes amortize the constant (problem-size-independent) overheads of each NPU invocation more effectively.

To demonstrate the benefit of our minimal-reconfiguration approach, which requires updating only the shim cores and two runtime parameters to switch between problem sizes, we also compared it to a design that reconfigures the whole NPU array (one \textit{xclbin} configuration binary for each problem size).
On the first iteration of a new GEMM size, our approach is, on average, $3.5\times$ faster than reconfiguring the whole array.
On subsequent iterations of the same size, reconfiguration is no longer required, so the runtimes of both approaches are roughly identical.

\begin{figure}
    \includegraphics{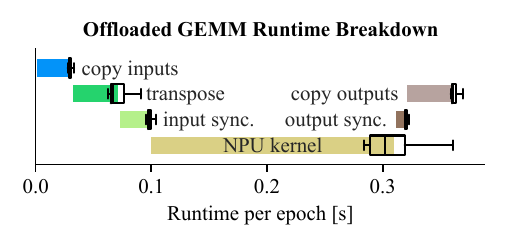}
    \caption{
        \label{fig:aie-details}
        Total runtime of all GEMM invocations of one training epoch, broken up by constituent stages (mean of 41 training epochs, lower is better).
        Our implementation copies input and output buffers from the GEMM call sites into XRT buffers for use with the NPU.
        Only some input matrices require transposition; where needed, the transpose also includes input copying.
        ``NPU kernel'' measures the actual GEMM being performed on the NPU.
        ``Input sync.'' and ``output sync.'' are unavoidable dispatch overheads incurred by the XDNA driver synchronizing CPU buffers with the NPU.
    }
\end{figure}

Figure \ref{fig:aie-details} shows the contributors to the overall runtime of all GEMM invocations.
Most of the time is spent running the NPU kernel;
however, CPU-side preparation work (copying, transposing, and synchronizing buffers) is also a significant contributor.

We also evaluated the numerical accuracy of our GEMM algorithm.
Our NPU kernel consumes \textit{bfloat16} inputs and accumulates and outputs \textit{float32} values, whereas the original CPU implementation operates on \textit{float32} values for both inputs and outputs.
Porting the CPU implementation to also use \textit{bfloat16} would slow it down significantly. The \textit{float32}-based CPU-implementation lowers to highly efficient vector FMA instructions on the CPU, (e.g. \texttt{vfmadd213ps}),
but such instructions do not exist for \textit{bfloat16}.
Therefore, our baseline is the unmodified CPU implementation that uses \textit{float32} inputs.
This leads to small numerical divergences between the CPU and NPU implementations; 
however, despite using a lower-precision data type, the NPU implementation achieves a slightly better validation error after 41 epochs.
Overall, the NPU output is numerically close to the CPU reference: the mean relative divergence is below $0.06\%$ (standard deviation $0.03\%$). The maximum deviation from the reference occurs for the $50304\times 256 \times 768$ size and is $0.1\%$.

\subsection{End-to-end Results}

\begin{figure}
    \includegraphics{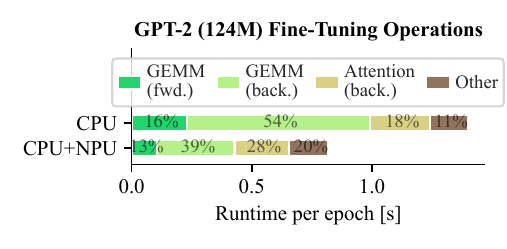}
    \caption{
        \label{fig:op-breakdown}
        Total runtime per training epoch of GPT-2 small (124M) fine-tuning with \textit{llm.c} in both the vanilla version (CPU) and our version with offloaded GEMMs (CPU+NPU), split by major constituent operations (mean across 41 training epochs, lower is better).
    }
\end{figure}

By plugging our optimized GEMM implementation into \textit{llm.c}, we can improve end-to-end performance of this application.
Figure \ref{fig:op-breakdown} shows which individual operations in \textit{llm.c} contribute to overall runtime.

Evidently, matrix multiplication operations dominate overall runtime, making them a worthwhile target for offloading to the NPU.
Thanks to the unified L3 memory, runtimes of the other unaltered operations on the CPU remain the same.

\begin{figure}
    \begin{center}
        \textbf{End-to-end Application Performance}
    \end{center}
    \includegraphics{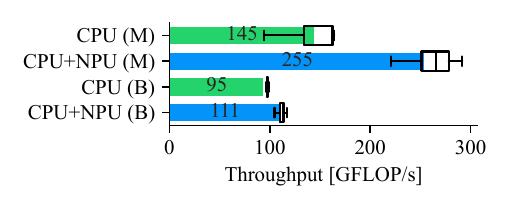}
    \includegraphics{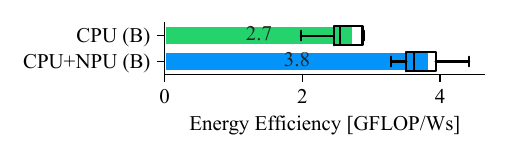}
    \caption{
        \label{fig:end-to-end}
        Throughputs (higher is better) of the vanilla \textit{llm.c} implementation of GPT-2 (CPU) and our version with GEMMs offloaded (CPU+NPU).
        ``(M)'' refers to benchmarks run on mains power, ``(B)'' to benchmarks run on battery power. 
        The improvements stem entirely from the offloaded GEMM operations; the remainder of computation still occurs on the CPU.
    }
\end{figure}

Offloading yields appreciable speedups in end-to-end throughput and energy efficiency, as figure \ref{fig:end-to-end} shows.
The reduced runtime and lower power consumption of the NPU-based implementation compound in the combined throughput-per-Watt-second metric, giving the NPU-based implementation an 
$\mathbf{1.4\times}$ 
edge over the CPU-based implementation.
Raw throughput is improved as well, at a speedup of $\mathbf{1.7\times}$ and $\mathbf{1.2\times}$ for mains and battery power, respectively.

%% file: future_work.tex
\section{Discussion and Future Work}

Our implementation shows that performance and power efficiency gains can be achieved by swapping in an NPU-based implementation for the most compute-intensive operations.
Our low-level approach enables us to tailor our NPU design to the application at hand (e.g., by choosing appropriate matrix tiling sizes and limiting ourselves to the minimal functionality required by the application) and avoid overheads that a ``one-size-fits-all'' framework may accumulate (such as matrix padding).
Hardware-level NPU programming harmonizes well with the central motif of \textit{llm.c}, which trades established ML frameworks for a bespoke pure-C implementation.

Although our individual matrix multiplication speedups and end-to-end gains are significant, a comparison with the advertised hardware capabilities shows that there is still headroom for future optimizations.
The theoretical compute capabilities of the NPU are on the order of \textit{tera-}FLOP/s, whereas the end-to-end throughput of our application is on the order of hundreds of \textit{giga-}FLOP/s only.
The reason for this disparity is twofold:
First, since our implementation still executes all but the most intensive operations on the CPU, execution performance quickly becomes constrained by the CPU's capabilities.
Second, our evaluation (figure \ref{fig:aie-details}) identifies the CPU-to-NPU-to-CPU data round trip at each kernel invocation as a costly overhead.

Tantalizingly, our results indicate that providing a bare-metal programming toolchain is the best path towards harnessing the full power of the available hardware, as it enables the straightforward investment of additional programming effort to implement the entire computation pipeline as an NPU design (thus eliminating CPU computation and the aforementioned CPU-to-NPU-to-CPU data movement bottlenecks). 
Other work has demonstrated the feasibility of such data-flow approaches.

In the near term, keeping with our approach of offloading only individual compute-intensive operations, 
future work may address some of the shortcomings of our current implementation,
for example by implementing zero-copy buffers for inputs and outputs and eliminating the need for transpose operations on the CPU through a more substantial rewrite of \textit{llm.c}.

%% file: acknowledgments.tex
\section{Acknowledgments}

Thank you to Joseph Melber, Kristof Denolf, and Phil James-Roxby from Advanced Micro Devices Inc.,
for their guidance and help in developing the NPU design and contributions to this research.